# Third Harmonic Generation at 223nm in the Metallic Regime of GaP

V. Roppo[1,2], J. Foreman[2] N. Akozbek[2], M.A. Vincenti[3], and M. Scalora[2]

[1]Universitat Politècnica de Catalunya, Departament de Física i Enginyeria Nuclear, Colom 11, 08222 Terrassa, Spain

[2]Charles M. Bowden Research Center AMSRD-AMR-WS-ST, RDECOM, Redstone Arsenal, Alabama 35898-5000, USA

[3] AEgis Technologies Group, 410 Jan Davis Drive, Huntsville, Alabama 35806, USA

**Abstract**

We demonstrate second and third harmonic generation from a GaP substrate 500μm thick. The second harmonic field is tuned at the absorption resonance at 335nm, and the third harmonic signal is tuned at 223nm, in a range where the dielectric function is negative. These results show that a phase locking mechanism that triggers transparency at the harmonic wavelengths persists regardless of the dispersive properties of the medium, and that the fields propagate hundreds of microns without being absorbed even when the harmonics are tuned to portions of the spectrum that display metallic behavior.

The solutions of the Maxwell's equations in nonlinear dielectrics that display second harmonic generation (SHG) have been thoroughly studied since the early work of Bloembergen and Pershan [1]. It was shown that in the undepleted pump regime the solution is composed of two parts: one resolves the homogeneous equation, while the other is a particular solution of the inhomogeneous equations driven by the nonlinear polarization term. For the case of SHG the homogenous component has traditionally been considered to be the only important constituent and so investigations have focused on it. Nevertheless, observations of a double-peaked SH signal in the strong phase mismatched regime have been reported (see for example [2,3]). Worthy of note is the work of Mlejnek et al. [4] where the authors showed the predictions made in reference [1], i.e. that the inhomogeneous component of the second harmonic (INH-SH) signal travels at the group velocity of the pump pulse. In a later study phase and group velocity matching were clearly demonstrated in Lithium Niobate in the range of transparency for all fields. The INH-SH was shown to refract at the same angle, and travel at the same velocity, as the pump [5]. At the same time, the homogeneous component of the SH refracts and travels according to the values one expects from material dispersion at that frequency.

The INH-SH component has not attracted much attention because the main goal has been and remains the achievement of efficient harmonic generation for application purposes. The inhomogeneous component is generally difficult to observe because it travels locked under the pump pulse, with conversion efficiencies that for bulk materials are still relatively low. Nevertheless, in a recent study results were extended and generalized to include third harmonic generation (THG) in both positive and negative index media [6]. In general, when a pump signal traverses an interface into a nonlinear medium it generates SH and/or TH fields, depending on the type of nonlinearity involved. Each harmonic consists of two components: (i) a homogeneous portion that eventually walk-off from the pump field; (ii) an inhomogeneous component that will always be phase and travels locked to the pump, with no energy transfer between the fields. The key observation here was the fact that the INH-SH has a k-vector always double that of the pump, even for large phase mismatches. Thus, assuming an incident pulse of arbitrary duration, the trapped SH signal has a carrier frequency that oscillates at $2\omega$ and a carrier wave-vector $2k_\omega$. It is then straightforward to conclude that the inhomogeneous SH signal experiences an *effective complex index* of refraction given by $2k_\omega c/2\omega = k_\omega c/\omega = n_\omega$, i.e. the same index of refraction as the pump pulse.

To experimentally prove this idea, in [7] a pump pulse tuned at 1300nm was launched into a slab of GaAs 500μm thick. Transmitted SH and TH signals were detected at 650 and 433nm respectively, far below the material absorption band edge (~900nm). Simulations showed that the homogeneous SH and TH signals were quickly absorbed, while the inhomogeneous components continued on through the sample experiencing an effective complex index of refraction identical to the index of the pump, which is tuned in a region of transparency. These results confirmed that the two components of the generated harmonics follow different destinies. A detailed investigation of the propagation characteristics of the inhomogeneous component in different circumstances can bring about new interesting scenarios. For example, reference [8] reports experimental evidence that the INH-SH can be enhanced by several orders of magnitude compared to bulk [7] in an opaque GaAs cavity environment, raising some expectation that the conversion efficiency could be significantly improved further. As also shown in reference [9], this occurs thanks to the double action of the synchronized pump and INH-SH field localization and consequent good overlap between the fields. Moreover, the authors in [10] suggest that the values of the effective nonlinearity of semiconductors in the opaque region of the spectrum can reach values between 500 and 1000pm/V.

Additional experimental evidence reported in reference [11] suggests that in a high-Q GaAs cavity the INH-SH signal (612nm) can achieve conversion efficiencies of the order of $10^{-3}$ with pumping field intensities as low as 0.15MW/cm$^2$ inside the cavity. Following the evolution of this topic, it now comes natural to ask the following question: does this phenomenon hold for harmonic fields tuned at frequencies in the metallic range? That is to say, if the pump field is tuned in the transparency region, will a harmonic field be able to propagate if it happens to be tuned in a region where $sign(\varepsilon) \neq sign(\mu)$, where one expects no propagating solutions? The short answer to this question is *yes*. In what follows we provide experimental and numerical evidence that show how an electromagnetic field is generated and propagates in otherwise forbidden wavelength ranges. In Fig.1 we show the dispersion data of GaP as given in reference [12]. The pump pulse is tuned at ~670nm, in the transparency region of the material, where $\varepsilon_{670nm} \approx 10.7$. Consequently, the SH field falls at the UV absorption resonance located near 335nm, where $\varepsilon_{335nm} = 19.1 + i\,24.1$. In turn, the TH signal is tuned to 223nm, where $\varepsilon_{223nm} = -9.6 + i\,10.4$ (metallic range). With these dispersion values, two externally incident fields tuned at or below 335nm would be completely reflected and/or absorbed within a few nanometers from the surface.

The experimental setup used to measure transmitted harmonics in GaP is shown in Fig.1. Pump pulses 80fs in duration (pulse waist) centered at ~670nm were generated by an optical parametric amplifier and amplified Ti:sapphire laser system operating at 1kHz repetition rate. Prior to sending the pump beam into the experimental setup, an equilateral dispersing prism and spatial filtering were used to remove background harmonics present in the beam path. A color filter was placed directly before the sample to remove any residual harmonics. At the sample position, the collimated pump beam's pulse energy was 2.7mJ, corresponding to a peak intensity of approximately 480MW/cm$^2$. After passing through the sample, the residual pump was filtered and the harmonics were highly reflected off by two set of four long wave pass (LWP) dichroic beam splitters. The second (third) harmonic mirrors each transmitted >90% of the pump and reflected >97% (>95%) of the harmonic. While propagating through the sequence of LWP mirrors, a single polarization of harmonic light was selected by a polarizer with single-pass extinction ratio of ~10$^{-3}$. Each harmonic was reflected back through the mirrors and polarizer by a concave aluminum mirror with deep-ultraviolet coating. Therefore, the residual pump light was suppressed by at least eight orders of magnitude and either the TM or TE component of the harmonic was selected with a contrast of at least 10$^6$. After passing back through the mirrors and polarizer, the harmonic was focused into an all-silica fiber coupled to a UV-sensitive spectrometer and liquid nitrogen cooled CCD.

In Fig.2 we show the measured spectra (solid curves) for the case of a TM polarized pump field. The <100> GaP 500μm thick wafer was illuminated at an angle of 40°. The basic results may be summarized as follows: the harmonic conversion efficiencies for the transmitted fields are approximately $10^{-9}$ for the TE-polarized SH at 335nm, and $\sim 1.8 \times 10^{-12}$ for the TM-polarized TH at 223nm. The detection limit of the system corresponds to a minimum detectable efficiency of $\sim 10^{-14}$. Any background harmonic present with the sample removed was below this detection limit.

We obtained theoretical confirmation of our experimental results by numerically solving Maxwell's equations in the time domain. The material was modeled as a collection of doubly resonant harmonic oscillators (see Fig.1) with a nonlinear polarization of the usual type: $\mathbf{P}_{NL} = \chi^{(2)}\mathbf{E}^2 + \chi^{(3)}\mathbf{E}^3$. The model is described in details in reference [13]. Suffice it to say here that the oscillator model is exemplified by the following scaled equations of motion that describe generic field envelope functions that are allowed to vary rapidly in space and time and is valid for pump depletion [13]:

$$\ddot{\mathbf{P}}_\omega + \tilde{\gamma}_\omega \dot{\mathbf{P}}_{b,\omega} + \tilde{\omega}_{0,\omega}^2 \mathbf{P}_{b,\omega} \approx \frac{n_0 e^2 \lambda_0^2}{m^* c^2} \mathbf{E}_\omega + \frac{e \lambda_0}{m^* c^2} \begin{pmatrix} -\frac{1}{2}\mathbf{E}_\omega^* \nabla \cdot \mathbf{P}_{2\omega} \\ +2\mathbf{E}_{2\omega} \nabla \cdot \mathbf{P}_\omega^* \\ -\frac{2}{3}\mathbf{E}_{2\omega}^* \nabla \cdot \mathbf{P}_{3\omega} \\ -\frac{3}{2}\mathbf{E}_{3\omega} \nabla \cdot \mathbf{P}_{2\omega}^* \end{pmatrix} + \frac{e \lambda_0}{m^* c^2} \begin{pmatrix} \left(\dot{\mathbf{P}}_\omega^* + i\omega\mathbf{P}_\omega^*\right) \times \mathbf{H}_{2\omega} \\ +\left(\dot{\mathbf{P}}_{2\omega} - 2i\omega\mathbf{P}_{2\omega}\right) \times \mathbf{H}_\omega^* \\ +\left(\dot{\mathbf{P}}_{2\omega}^* + 2i\omega\mathbf{P}_{2\omega}^*\right) \times \mathbf{H}_{3\omega} \\ +\left(\dot{\mathbf{P}}_{3\omega} - 3i\omega\mathbf{P}_{3\omega}\right) \times \mathbf{H}_{2\omega}^* \end{pmatrix}, (1)$$

$$\ddot{\mathbf{P}}_{2\omega} + \tilde{\gamma}_{2\omega} \dot{\mathbf{P}}_{2\omega} + \tilde{\omega}_{0,2\omega}^2 \mathbf{P}_{2\omega} \approx \frac{n_0 e^2 \lambda_0^2}{m^* c^2} \mathbf{E}_{2\omega} + \frac{e \lambda_0}{m^* c^2} \begin{pmatrix} \mathbf{E}_\omega \nabla \cdot \mathbf{P}_\omega \\ -\frac{1}{3}\mathbf{E}_\omega^* \nabla \cdot \mathbf{P}_{3\omega} \\ -3\mathbf{E}_{3\omega} \nabla \cdot \mathbf{P}_\omega^* \end{pmatrix} + \frac{e \lambda_0}{m^* c^2} \begin{pmatrix} \left(\dot{\mathbf{P}}_\omega - i\omega\mathbf{P}_\omega\right) \times \mathbf{H}_\omega \\ +\left(\dot{\mathbf{P}}_\omega^* + i\omega\mathbf{P}_\omega^*\right) \times \mathbf{H}_{3\omega} \\ +\left(\dot{\mathbf{P}}_{3\omega} - 3i\omega\mathbf{P}_{3\omega}\right) \times \mathbf{H}_\omega^* \end{pmatrix}, (2)$$

$$\ddot{\mathbf{P}}_{3\omega} + \tilde{\gamma}_{3\omega} \dot{\mathbf{P}}_{3\omega} + \tilde{\omega}_{0,3\omega}^2 \mathbf{P}_{3\omega} \approx \frac{n_0 e^2 \lambda_0^2}{m^* c^2} \mathbf{E}_{3\omega} + \frac{e \lambda_0}{m^* c^2} \begin{pmatrix} \frac{1}{2}\mathbf{E}_\omega \nabla \cdot \mathbf{P}_{2\omega} \\ +2\mathbf{E}_{2\omega} \nabla \cdot \mathbf{P}_\omega \end{pmatrix} + \frac{e \lambda_0}{m^* c^2} \begin{pmatrix} \left(\dot{\mathbf{P}}_{2\omega} - 2i\omega\mathbf{P}_{2\omega}\right) \times \mathbf{H}_\omega \\ +\left(\dot{\mathbf{P}}_\omega - i\omega\mathbf{P}_\omega\right) \times \mathbf{H}_{2\omega} \end{pmatrix}. (3)$$

In Eqs.(1-3) the spatial derivatives ($\nabla \cdot$ operator) are performed with respect to scaled longitudinal and transverse coordinates, $\xi = z/\lambda_0$ and $\tilde{y} = y/\lambda_0$ respectively, where $\lambda_0 = 1\mu m$ is arbitrarily chosen as the reference wavelength. We also scale time as $\tau = ct/\lambda_0$. $e$ and $m^*$ are the electron charge and effective mass, respectively. For optical purpose, typical effective electron masses are of order $m^* \sim 0.05 m_e$ [14]. In our calculations we use $m^* \sim 0.025 m_e$. $\mathbf{P}_{N\omega}$ is the polarization, $\mathbf{E}_{N\omega}$ and $\mathbf{H}_{N\omega}$ are the electric and the magnetic fields associated with the $N^{th}$ harmonic. As in reference [13] we also operate in a two-dimensional space, and allow the simultaneous presence of TE- and TM-polarized fields so that each has a two-dimensional spatial description. The scaled coefficients are $\tilde{\gamma}_{N\omega} = (\gamma - Ni\omega)$, the damping coefficient, and $\tilde{\omega}_{0,N\omega}^2 = (\omega_0^2 - (N\omega)^2 + i\gamma N\omega)$, the resonance frequency, where $N$ is an integer that denotes the given harmonic order. This theoretical description takes into account harmonic generation that arises from all interface crossings (spatial derivatives on the polarization vectors) and from the ever-present magnetic Lorentz force, which is usually neglected outside the context of metals but that nevertheless contributes to the process and is included here for completeness. The total polarization that is fed back into Maxwell's equation is the vector sum of each of the solutions of Eqs.(1-3) and the nonlinear polarization that arises from $\chi^{(2)}$ and $\chi^{(3)}$ phenomena.

When the pump is TM-polarized the intrinsic nature of the $\chi^{(2)}$ tensor of GaP (the only non-zero components are $d_{14}=d_{25}=d_{36}$) selects the TE-polarized SH signal. The simulations show that the homogeneous component is completely absorbed within just a few nanometers from the entry surface, leaving only the INH-SH. Harmonic signals generated at the surface due to symmetry breaking that inevitably take place during the interaction, on the other hand, have the same polarization as the incident pump pulse [13], but have much smaller conversion efficiencies compared to $\chi^{(2)}$ contributions. Thus, cross checking the polarization of the generated signal can be used to determine its origin, i.e. surface or bulk. Using the theoretical approach outlined in reference [13] we determined that the efficiency of the transmitted, surface- generated, TM-polarized SH field is of order $10^{-11}$, i.e. at least two orders of magnitude smaller than the TE-polarized component that arises from the intrinsic $\chi^{(2)}$ tensor of the medium. The experiment confirms these findings. As a result, there is little doubt that most of the TE-polarized signal originates with the bulk $\chi^{(2)}$ of GaP.

For THG, nonlinear surface and volume sources that arise from Coulomb and magnetic Lorentz forces only [13] are inadequate to generate an observed transmitted, TM-polarized TH signal with an efficiency of $10^{-12}$. Our calculations suggest that integrating Maxwell's equations coupled to Eqs.(1-3) above with the exclusion of bulk $\chi^{(2)}$ and $\chi^{(3)}$ contributions yields TH conversion efficiencies of order $10^{-20}$. An adequate prediction of THG can be made by adopting the realistic circumstances of a third order $\chi^{(3)}$ tensor of cubic type, with $\bar{4}3m$ symmetry and four independent components [15], namely $\chi^{(3)}_{xxxx}$, $\chi^{(3)}_{yyxx}$, $\chi^{(3)}_{yzyz}$, $\chi^{(3)}_{yzzy}$, with $\chi^{(3)}_{xxxx} = \chi^{(3)}_{yyyy} = \chi^{(3)}_{zzzz}$. The remaining symmetry properties of the tensor for this class of materials may be consulted in reference [15]. Our model thus includes surface and volume contributions from Coulomb and magnetic Lorentz forces, as well as contributions from the realistic nonlinear second and third order tensors that give rise to SHG and THG.

In Fig.3 we show the results as a superposition of different temporal snapshots of the incident and generated harmonic pulses. The interaction proceeds as follows. As the input pulse (Fig.3a) enters the material SHG and THG occur. In the figure we report the most intense components, namely TE-polarized SH (Fig.3b) and TM-polarized TH (Fig.3c) pulses. The homogeneous components quickly vanish due to either the high absorption (SH) or the metallic environment (TH). The inhomogeneous components are the only portions of the harmonics that survive, and as can be clearly seen from the figure, they perfectly overlap with the pump pulse at all times. The amount of energy transferred to each harmonic by the pump pulse is dictated by the material index mismatch at the interface, and does not vary during the propagation [6]. When the pump pulse reaches the exit surface the fields become uncoupled and are able to propagate freely. In Fig.3d we also show the transmitted field energies by monitoring them as a function of time (bottom right).

The calculated spectra are shown in Fig.2 (dashed lines) and they overlap well the experimental data when we choose $\chi^{(2)} = 2d_{14} = 2d_{25} = 2d_{36} \approx 500\, pm/V$, and $\chi^{(3)}_{xxxx} = \chi^{(3)}_{yyyy} = \chi^{(3)}_{zzzz} \approx 4 \times 10^{-19}\, m^2/V^2$. Although the remaining third order tensor components $\chi^{(3)}_{yyxx}$, $\chi^{(3)}_{yzyz}$, and $\chi^{(3)}_{yzzy}$ are also included in the calculation and are taken to be of the same order as the diagonal terms, they contribute little the overall TH conversion efficiency. The input parameters we used in our simulations closely reflect those of the experiment. Our calculations thus suggest that indeed the second order coefficients tend to achieve relatively large values in the UV range, as also reported elsewhere [10]. In contrast, the magnitude of the third order coefficient is more in line with typical reported values.

Beside some aspects that clearly appear to have fundamental relevance, the discovery that the phase locking phenomenon still applies in the UV regime for harmonics tuned in spectral ranges where the medium displays metallic behavior may be of some technological importance in re-thinking and re-designing semiconductor-based integrated devices at the nanoscale. Harmonic generation is only one aspect of the role that common semiconductors can play at UV wavelengths, i.e. ranges where typical semiconductors distinguish themselves only for their opacity. As another example, semiconductor-based super-lenses [16] and enhanced transmission gratings [17] that operate in the UV range have already been proposed. Taken together, these results show that the functionality of semiconductors may be pushed down the wavelength scale, toward the soft-Xray region of the spectrum. The natural next step to take beyond this work is to study the dynamics in cavity environments, as was done previously [8, 11], possibly in metal-semiconductor structures [13], where the $\chi^{(3)}$ of metals can be fully exploited with the clear aim to achieve high conversion efficiencies in the deep UV range. In summary, we have experimentally and theoretically demonstrated the propagation of light pulses in the ultra-violet metallic frequency range of GaP. The pulse was generated near the entry surface of the sample as a third harmonic signal using quadratic and cubic nonlinearities of the material. The propagation was possible thanks to a phase locking mechanism that binds inhomogeneous components to the fundamental field, an effect that persists regardless of the dispersive characteristics of the medium at the harmonic wavelengths.

**Acknowledgement**

V.R. acknowledges partial financial support from the Army Research Office (W911NF-10-2-0105). N.A. acknowledges financial support provided by the National Research Council.

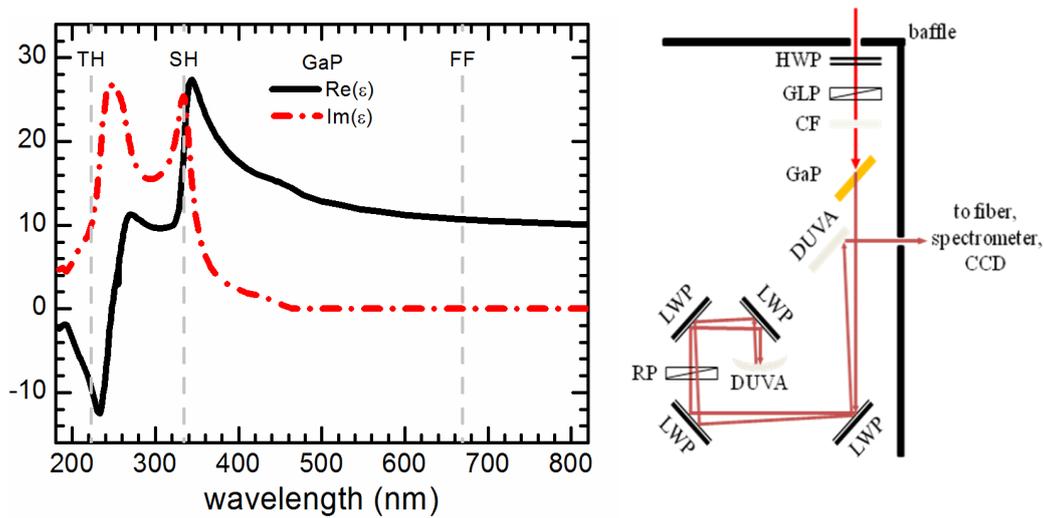

**Figure 1**: **(Left)** dispersion relation of GaP from reference [12]. The dashed lines indicate tuning of the relative harmonic fields. Two absortpion resonances can be identified, at ~330nm and ~240nm. **(Right)** Experimental Set-up. The polarization of input pump pulses at 670 nm is established by a half-wave plate (HWP) and Glan-laser polarizer (GLP). A color filter (CF, Corning 3-73) removes the harmonics before the pumps strikes the GaP sample. The harmonics reflect off four long wave pass dichroic beam splitters (LWP). A Rochon polarizer (RP) transmits only one polarization of harmonic light. A concave aluminum mirror with deep-ultraviolet coating (DUVA) retro-reflects the harmonic back through the RP/LWP system and focuses the harmonic into an all-silica fiber for detection

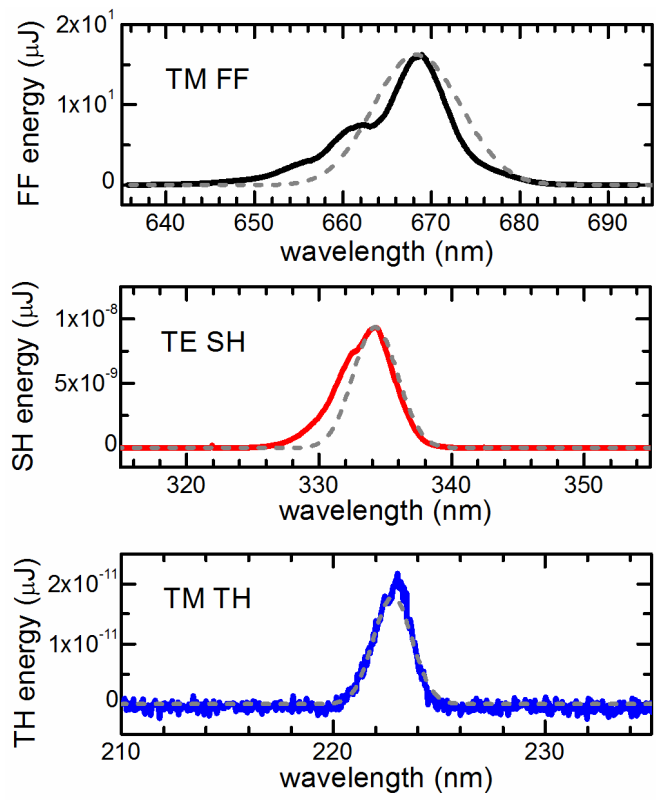

**Figure 2**: Measured spectra for the TM-polarized pump (continuous lines). Results of the numerical simulation (dashed lines).

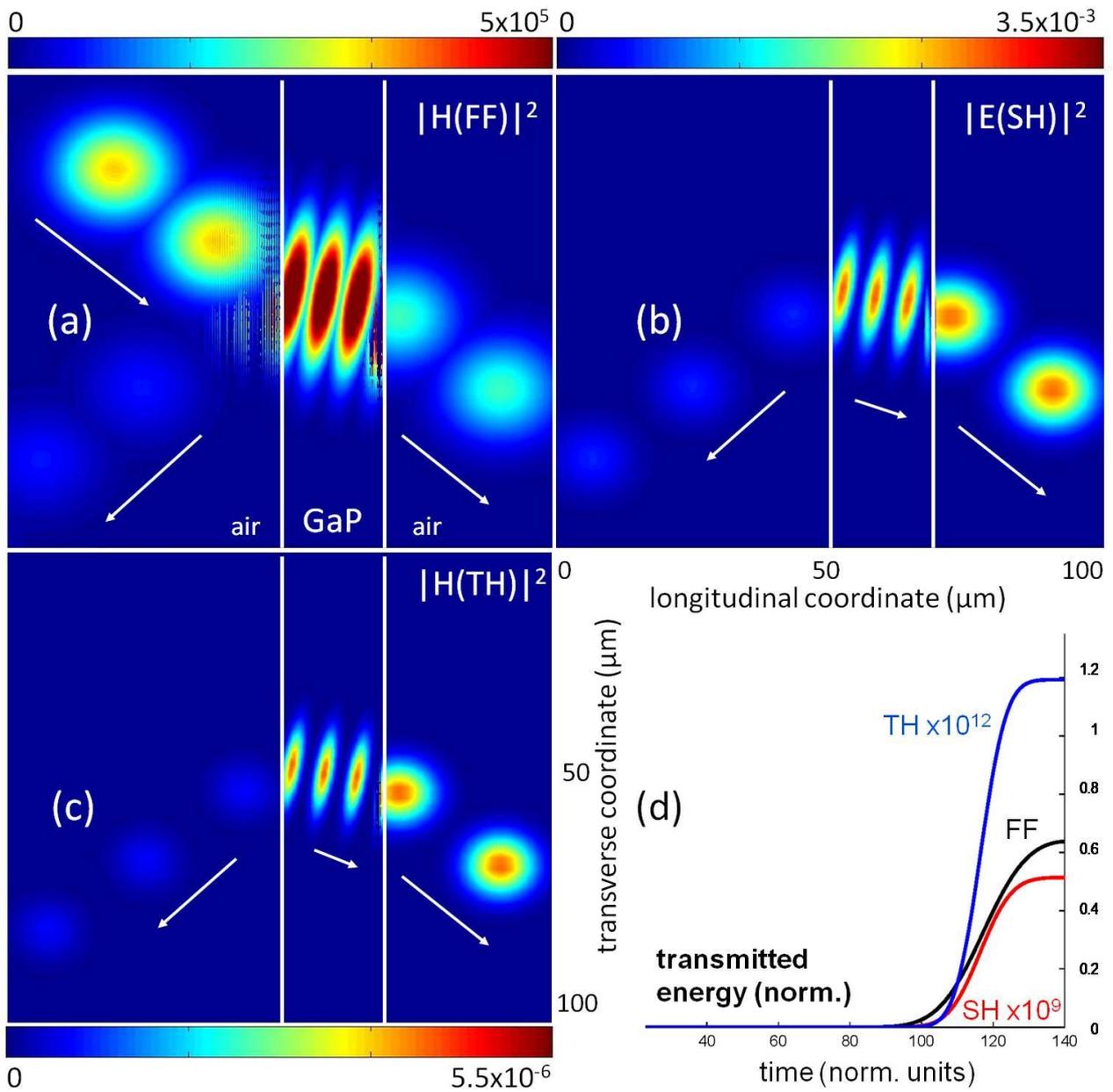

**Figure 3**: Time propagation calculation that shows superposition of different temporal snapshots. An 80fs pump pulse impinges at 40º on a GaP 20μm-thick slab (a) and SH (b) and TH (c) signals are generated. The harmonic signals keep propagating inside the bulk material perfectly overlapped to the pump pulse without been absorbed, quite insensitively to substrate thickness. (d): Field energies as a function of time at the right of the sample.